\documentclass{article}

 \usepackage[final,nonatbib]{nips_2018}
\usepackage[superscript]{cite}
\usepackage{graphicx,longtable,rotating}

\usepackage{amsmath,hyperref}
\usepackage[utf8]{inputenc} 
\usepackage[T1]{fontenc}    
\usepackage{hyperref}       
\usepackage{url}            
\usepackage{booktabs}       
\usepackage{amsfonts}       
\usepackage{nicefrac}       
\usepackage{microtype}      

\def\ff{\frac}

\def\aa{\alpha}

\def\ss{\sigma^2}

\def\bal{\begin{align}}

\def\cs{\chi^2}

\def\gcp{\text{gcp}}
\def\gcphat{\hat{\text{gcp}}}

\title{Distinguishing correlation from causation using genome-wide association studies}

\author{
  Luke~J.~O'Connor \\
 Department of Epidemiology\\
Harvard T.H. Chan School of Public Health\\
 Boston, MA 02115 \\
  \texttt{loocnnor@g.harvard.edu} \\
\And
 Alkes~L.~Price\\
 Department of Epidemiology\\
Harvard T.H. Chan School of Public Health\\
 Boston, MA 02115 \\
  \texttt{aprice@hsph.harvard.edu}
  }

\begin{document}

\maketitle

\begin{abstract}
Genome-wide association studies (GWAS) have emerged as a rich source of genetic clues into disease biology, and they have revealed strong genetic correlations among many diseases and traits. Some of these genetic correlations may reflect causal relationships. We developed a method to quantify causal relationships between genetically correlated traits using GWAS summary association statistics. In particular, our method quantifies what part of the genetic component of trait~1 is also causal for trait~2 using mixed fourth moments $E(\alpha_1^2\alpha_1\alpha_2)$ and $E(\alpha_2^2\alpha_1\alpha_2)$ of the bivariate effect size distribution. If trait~1 is causal for trait~2, then SNPs affecting trait~1 (large $\alpha_1^2$) will have correlated effects on trait~2 (large $\alpha_1\alpha_2$), but not vice versa. We validated this approach in extensive simulations. Across 52 traits (average $N=331$k), we identified 30 putative genetically causal relationships, many novel, including an effect of LDL cholesterol on decreased bone mineral density. More broadly, we demonstrate that it is possible to distinguish between genetic correlation and causation using genetic association data.
\end{abstract}
{\it This manuscript is an abridged version of O'Connor and Price 2018 Nature Genetics\cite{oconnor}}. 

Genome-wide association studies (GWAS) have identified thousands of common genetic variants (SNPs) affecting disease risk and other complex traits\cite{hapmap,purcell,okada,wood,visscher}. The same SNPs often affect multiple traits, resulting in a genetic correlation: genetic effect sizes are correlated across the genome, and so are the traits themselves \cite{bulik1,pickrell,pgc,verbanck}. Some genetic correlations may result from causal relationships. For example, SNPs that cause higher triglyceride levels reliably confer increased risk of coronary artery disease\cite{do}. This causal inference approach, using genetic variants as instrumental variables\cite{pearl,angrist}, is known as Mendelian Randomization (MR)\cite{smith0,voight,smith1}. However, genetic variants often have shared, or ``pleiotropic'', effects on multiple traits even in the absence of a causal relationship, and pleiotropy is a challenge for MR, especially when it leads to a strong genetic correlation \cite{smith0,smith1,pickrell,bulik1,verbanck}. Statistical methods have been used to account for certain kinds of pleiotropy\cite{bowden,verbanck,pickrell,bowden1,kang,hartwig}; however, these approaches too are easily confounded by genetic correlations due to pleiotropy. Here, we develop a robust method to distinguish whether a genetic correlation results from pleiotropy or from causality.

\begin{figure}[h]
\includegraphics[width=1.1\textwidth]{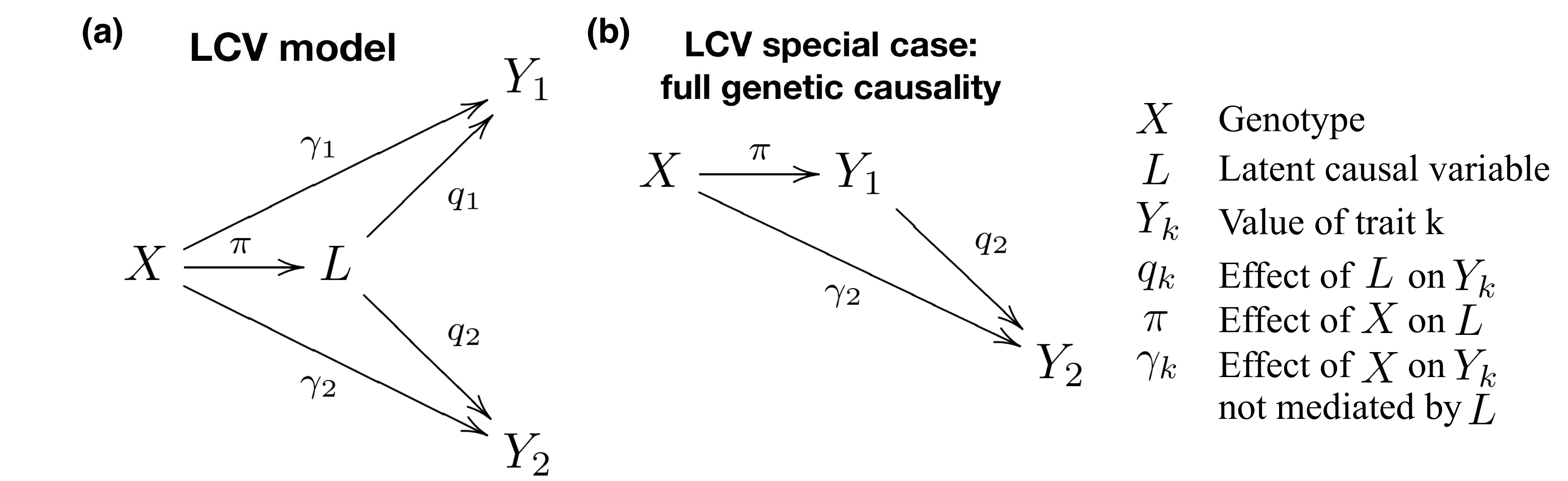}
\caption{Illustration of the latent causal variable model. We display the relationship between genotypes $X$, latent causal variable $L$ and trait values $Y_1$ and $Y_2$. (a) General case of the model. (b) Special case of full genetic causality: when $q_1=1$, the genetic component of $Y_1$ is equal to $L$. }\label{diagram-fig}
\end{figure}

\section{Latent causal variable model}
The latent causal variable (LCV) model features a latent variable $L$ that mediates the genetic correlation between the two traits (Figure~\ref{diagram-fig}a). More abstractly, $L$ represents the shared genetic component of both traits. Trait~1 is {\it fully genetically causal} for trait~2 if it is perfectly genetically correlated with $L$; ``fully'' means that the entire genetic component of trait~1 is causal for trait~2 (Figure~\ref{diagram-fig}b). More generally, trait~1 is {\it partially genetically causal} for trait~2 if the latent variable has a stronger genetic correlation with trait~1 than with trait~2; ``partially'' means that  part of the genetic component of trait~1 is causal for trait 2. In equations, $Y_1$ and $Y_2$ are modeled as linear functions of the genotype vector, $X$, and uncorrelated noise, $\epsilon$ (a linear model is appropriate, due to the small effect sizes of individual variants):
\bal\label{lcv-model}
Y_1=q_1X\pi+X\gamma_1+\epsilon_1,\quad Y_2=q_2X\pi+X\gamma_2+\epsilon_2.
\end{align}
where $q_k$ is a scalar representing the effect of $L$ on trait $k$, $X\pi$ represents the shared genetic component $L$, and $\gamma_kX$ represents the trait-specific genetic component of trait $k$ (see Figure~\ref{diagram-fig}). In order to quantify partial causality, we define the {\it genetic causality proportion} ($\gcp$) of trait~1 on trait~2. The gcp is defined as the number $x$ such that:
\begin{align}\label{gcp-def}
{q_2^2}/{q_1^2}=(\rho^2)^x,
\end{align}
where $\rho=q_1q_2$ is the genetic correlation\cite{bulik1}. The gcp ranges from $-1$ to $1$. When it is positive, $q_1$ is greater than $q_2$, and trait~1 is partially genetically causal for trait~2. When it is equal to 1, $q_1$ is equal to one, $Y_1$ is fully genetically correlated with $L$, and trait~1 is fully genetically causal. When it is 0, there is no partial causality; $L$ explains the same proportion of heritability for both traits. A high value of $\gcp$ implies that interventions targeting trait~1 are likely to affect trait~2, and an intermediate value implies that some interventions targeting trait 1 may affect trait 2, depending on their mechanism of action. However, we caution that an intervention may fail to mimic genetic perturbations, e.g. due to its timing relative to disease progression.

The LCV model makes one fundamental assumption, that the bivariate effect size distribution $(\aa_1,\aa_2)$ is a sum of two independent distributions: (1) a shared genetic component $(q_1\pi, q_2\pi)$, whose values are proportional for both traits; and (2) a trait-specific distribution $(\gamma_1,\gamma_2)$ whose density is mirror symmetric across both axes (i.e.,
$(\gamma_1,\gamma_2)\sim(-\gamma_1,\gamma_2)\sim(\gamma_1,-\gamma_2)$). 
This assumption is much weaker than the ``exclusion restriction" assumption of MR\cite{smith1}; in particular, the LCV model permits both correlated pleiotropic effects (mediated by $L$) and uncorrelated pleiotropic effects (not mediated by $L$), while the exclusion restriction assumption permits neither. Linkage disequilibrium (LD) is not central to the model, but this assumption pertains to the marginal effect size distribution (inclusive of LD). This model and the ways it could be violated are discussed in greater detail in the Appendix.

\section{Inference using fourth moments}
In order to test for partial genetic causality and to estimate the $\gcp$, we compare the mixed fourth moments $E(\alpha_1^2\alpha_1\alpha_2)$ and $E(\alpha_2^2\alpha_1\alpha_2)$ of the marginal SNP-effect-size distribution. The rationale for utilizing these mixed fourth moments is that if trait~1 is causal for trait~2, then SNPs with large effects on trait~1 (i.e. large $\aa_1^2$) will have proportional effects on trait~2 (large $\aa_1\aa_2$), so that $E(\alpha_1^2\alpha_1\alpha_2)$ will be large; conversely, SNPs with large effects on trait~2 (large $\aa_2^2$) will generally not affect trait~1 (small $\aa_1\aa_2$), so that $E(\alpha_2^2\alpha_1\alpha_2)$ will not be as large. Thus, estimates of the mixed fourth moments can be used to test for partial genetic causality and to estimate the $\gcp$.

In particular, we utilize the following relationship between the mixed fourth moments and the parameters $q_1$ and $q_2$ (Figure 1; see Appendix for derivation):
\bal\label{main-eq}
E(\alpha_1^3\alpha_2)=\kappa_\pi q_1^3q_2+3\rho,
\end{align}
where $\pi$ is the effect of a SNP on $L$ and $\kappa_\pi=E(\pi^4)-3$ is the excess kurtosis of $\pi$. This equation implies that if $E(\alpha_1^3\alpha_2)^2\geq E(\alpha_1\alpha_2^3)^2$, then $q_1^2\geq q_2^2$. We note that when $\kappa_\pi=0$, and in particular when $\pi$ follows a normal distribution, this equation is not useful for inference and the model is unidentifiable.

We calculate statistics $S(x)$ for each possible value of $\gcp=x$, based on equation \eqref{main-eq}. These statistics utilize estimates of the heritability and the genetic correlation\cite{bulik1,bulik2}, in addition to the mixed fourth moments. We estimate the variance of these statistics using a block jackknife and obtain an approximate likelihood function for the $\gcp$. We compute a posterior mean estimate of $\gcp$ (and a posterior standard deviation) using a uniform prior on $[-1,1]$. We test the null hypothesis (that $\gcp=0$) using the statistic $S(0)$. Further details of the method are provided in the Appendix. Open source implementations of the LCV method in Matlab and R are available at \url{github.com/lukejoconnor/LCV}.
\begin{figure}[h]
\includegraphics[width=\textwidth]{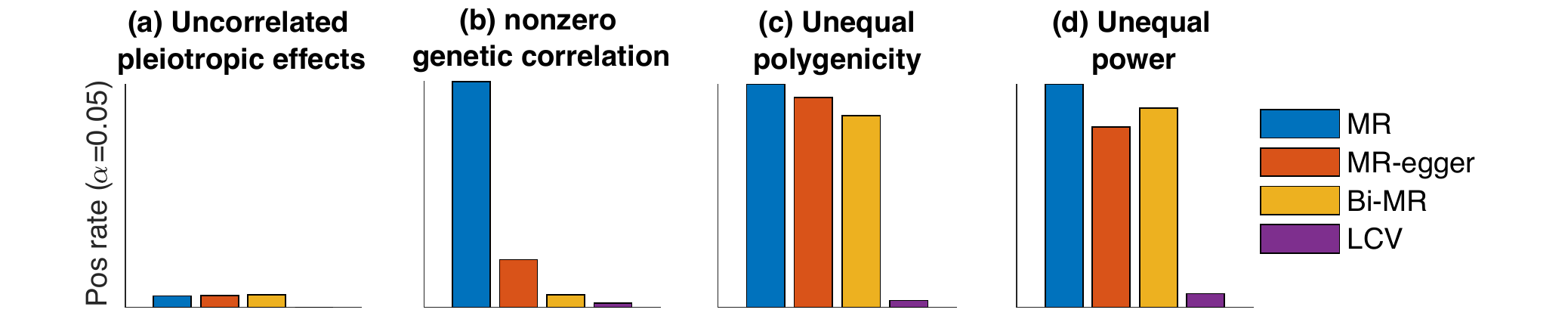}
\caption{Null simulations with no LD to assess calibration. We report the false positive rate ($\alpha=0.05$) for a causal (or partially causal) effect. (a) Pleiotropic SNPs have uncorrelated effects on the two traits ($\rho=0$). (b) Pleiotropic SNPs have correlated effects on the two traits, resulting in a genetic correlation ($\rho=0.2$). (c) Trait-specific SNPs have unequal polygenicity ($4\times$ different) between the two traits ($\rho=0.2$). (d) The two traits have unequal sample size ($5\times$ different; $\rho=0.2$).}\label{MR-null-fig}
\end{figure}
\begin{figure}[h]
\includegraphics[width=\textwidth]{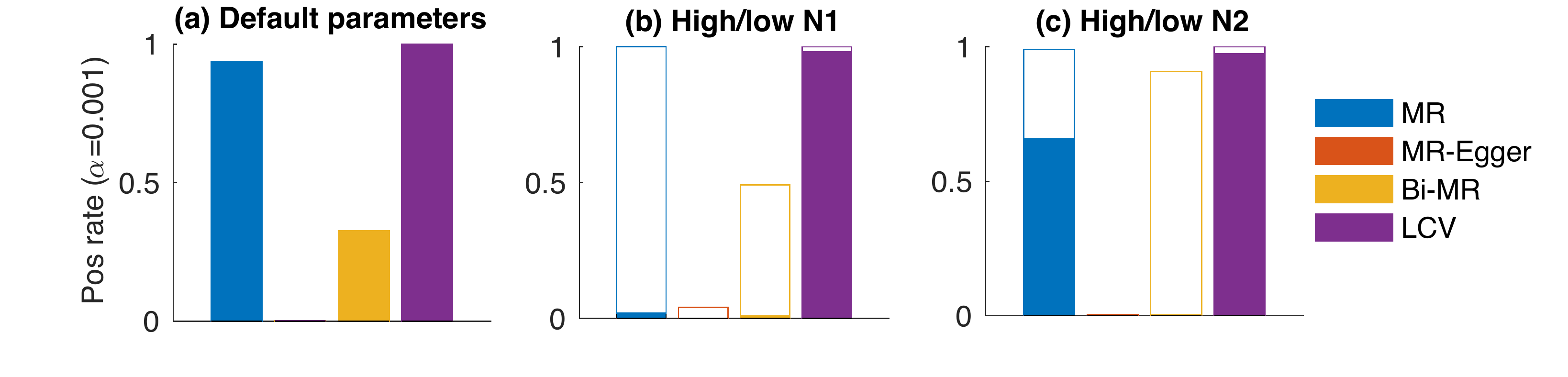}
\caption{Causal simulations with no LD to assess power. We report true the positive rate ($\alpha=0.001$) for a (partially) causal effect. (a) Causal simulations with default parameters. (b) Higher (unfilled) or lower (filled) sample size for trait~1. (c) Higher (unfilled) or lower (filled) sample size for trait~2.}\label{MR-power-fig}
\end{figure}
\section{Simulations}
To compare LCV with existing causal inference methods, we performed simulations involving simulated GWAS summary statistics with no LD. We compared four main methods: LCV, random-effect two-sample MR \cite{burgess0} (denoted MR), MR-Egger \cite{bowden}, and Bidirectional MR \cite{pickrell}. We applied each method to simulated GWAS summary statistics for two traits, testing for causality. 

We compared the false positive rate of LCV and MR  in null simulations ($\gcp=0$) with pleiotropy. There were $N=100$k samples, $M=50$k SNPs, and heritability $h^2=0.3$. First, we performed simulations with uncorrelated pleiotropic effects (via $\gamma_1,\gamma_2$; see Figure 1a) and zero genetic correlation. All four methods produced well-calibrated or conservative p-values (Figure~\ref{MR-null-fig}a). Second, we performed null simulations with a nonzero genetic correlation ($\rho=0.2$). MR and MR-Egger both exhibited severely inflated false positive rates; in contrast, Bidirectional MR and LCV produced well-calibrated p-values (Figure~\ref{MR-null-fig}b). Third, we performed null simulations with a nonzero genetic correlation and differential polygenicity in the non-shared genetic architecture between the two traits. Because causal SNPs affecting trait~1 only had larger effect sizes than SNPs affecting  trait~2 only, they were more likely to be genome-wide significant, and as a result, Bidirectional MR (and other MR methods) exhibited inflated false positive rates. In contrast, LCV produced well-calibrated p-values (Figure~\ref{MR-null-fig}c). Fourth, we performed null simulations with a nonzero genetic correlation and unequal sample size for the two traits, reducing the sample size from 100k to 20k for trait~2. The MR methods, including bidirectional MR, exhibited inflated false positive rates, unlike LCV (Figure~\ref{MR-null-fig}d).

We sought to compare the power of LCV and MR to detect a causal effect in simulations with full genetic causality ($\gcp=1$). First, we performed causal simulations with reduced sample size ($N=25$k), and a causal effect size of $q_2=0.2$. LCV and MR were well-powered to detect a causal effect, while Bidirectional MR and MR-Egger had lower power. 
Second, we reduced the sample size for trait~1 (Figure~\ref{MR-power-fig}b), finding that LCV had high power while the MR methods had very low power, owing to the small number of genome-wide significant SNPs. 
Third, we reduced the sample size for trait~2 (Figure~\ref{MR-power-fig}c). LCV and MR had high power, while Bidirectional MR and MR-Egger had lower power.

Simulations involving LCV model violations and LD are described in the Appendix. We conclude that MR methods, including methods that purportedly account for pleiotropy, are easily confounded. In contrast, LCV is well-calibrated and well-powered across diverse genetic architectures.

\section{Results on 52 diseases and complex traits}
We applied LCV and the MR methods to GWAS summary statistics for 52 diseases and complex traits, including 37 from UK Biobank\cite{sudlow,bycroft,loh} (average $N=337$k; see Table~\ref{phenotypes-table} in Appendix). 429 trait pairs (32\%) had a nominally significant genetic correlation ($p<0.05$). We applied LCV to these trait pairs and detected significant evidence of  full or partial genetic causality for 59 trait pairs (FDR $< 1\%$), including 30 trait pairs with $\gcphat>0.6$ (Table~\ref{all-sig-table} in Appendix).

Myocardial infarction (MI) had a nominally significant genetic correlation with 31 other traits, of which six had significant evidence for a fully or partially genetically causal effect on MI (Table~\ref{all-sig-table} in Appendix). Consistent with previous studies, these traits included LDL and high cholesterol \cite{cohen,voight}, triglycerides\cite{do} and BMI \cite{holmes}, but not HDL\cite{voight}. We also detected evidence for a fully or partially genetically causal effect of hypothyroidism; such an effect is mechanistically plausible \cite{klein,grais}.

We detected evidence for a negative genetically causal effect of LDL on bone mineral density (BMD; Table~\ref{all-sig-table}). This result is consistent with preliminary clinical evidence \cite{wang}, and familial defective apolipoprotein B leads to high LDL cholesterol and low bone mineral density \cite{yerges}. Larger clinical trials of LDL lowering for individuals at risk of osteoporosis may be warranted, although we caution that the effect size is not expected to be large.

In order to evaluate whether the limitations of MR observed in simulations are also observed in analyses of real traits, we applied MR to all 429 genetically correlated trait pairs. MR reported significant causal relationships (1\% FDR) for 271/429 trait pairs, including 155 pairs of traits for which each trait was reported to be causal for the other.  This implausible result confirms that MR frequently produces false positives due to genetic correlations, consistent with simulations (Figure~\ref{MR-null-fig}). 

Our method represents an advance for two main reasons. First, LCV reliably distinguishes between genetic correlation and causation. Unlike existing methods, LCV produces well-calibrated false positive rates in null simulations and plausible results on real data. Second, we define and estimate the genetic causality proportion (gcp) to quantify the degree of causality non-dichotomously. We found that 29/59 significant trait pairs had $\gcp$ estimates below 0.6; these trait pairs, while potentially interesting, probably do not reflect fully causal relationships. However, an important limitation of this study is that it models only two traits and only one latent variable; it cannot be used to perform conditional analyses with multiple traits, and it may miss causal effects when additional shared pathways also contribute to the genetic correlation.  Nonetheless, we anticipate that LCV will be widely used as genetic association data becomes available for increasing numbers of diseases and potentially causal traits.

\clearpage

{\bf \huge Appendix}
	\section{Tables}

\clearpage

{\small\centering
\begin{longtable}{c c c c c }
\hline
	 & Phenotype  & Reference & $N$ (thousands) & $Z_h$ \\ \hline
	 & Anorexia & Boraska et al., 2014 Mol Psych  & 32 & 17.8 \\ 
	 & Autism Spectrum & PGC Cross-Disorder Group, 2013 Lancet & 10 & 12.1 \\ 
	 & Bipolar Disorder & BIP Working Group of the PGC, 2011 Nat Genet  & 17 & 11.8 \\ 
	 & Breast Cancer & Amos et al., 2016 Cancer Epidemiol. Biomarkers Prev. & $\sim 447$* & 16 \\ 
	 & Celiac Disease & Dubois et al., 2010 Nat Genet & 15 & 10.4 \\ 
	 & Crohns Disease & Jostins et al., 2012 Nature  & 21 & 12.1 \\ 
	 & Depressive symptoms & Okbay et al., 2016 Nat Genet & 161 & 13.1 \\ 
	 & HDL & Teslovich et al., 2010 Nature & 98 & 8.2 \\ 
	 & HbA1c & Soranzo et al., 2010 Diabetes	 & 46 & 8.8 \\ 
	 & LDL & Teslovich et al., 2010 Nature  & 93 & 8.1 \\ 
	 & Lupus & Bentham et al., 2015 Nat Genet & 14 & 10.2 \\ 
	 & Prostate Cancer & Amos et al., 2016 Cancer Epidemiol. Biomarkers Prev. & $\sim 447$* & 7.5 \\ 
	 & Schizophrenia & SCZ Working Group of the PGC, 2014 Nature & 70 & 17.4 \\ 
	 & Triglycerides & Teslovich et al., 2010 Nature & 94 & 9.5 \\ 
	 & Ulcerative Colitis & Jostins et al., 2012 Nature  & 27 & 8.8 \\ 
	 & Eosinophil count & UK Biobank \citen{sudlow,bycroft,loh} & $\sim 460$** & 20.8 \\ 
	 & Reticulocyte count & UK Biobank \citen{sudlow,bycroft,loh} & $\sim 460$ & 19.9 \\ 
	 & Lymphocyte count & UK Biobank \citen{sudlow,bycroft,loh} & $\sim 460$ & 22.7 \\ 
	 & Mean corpuscular hemoglobin & UK Biobank \citen{sudlow,bycroft,loh} & $\sim 460$ & 14.3 \\ 
	 & Mean platelet volume & UK Biobank \citen{sudlow,bycroft,loh} & $\sim 460$ & 15.7 \\ 
	 & Monocyte count & UK Biobank \citen{sudlow,bycroft,loh} & $\sim 460$ & 15.1 \\ 
	 & Platelet count & UK Biobank \citen{sudlow,bycroft,loh} & $\sim 460$ & 20.2 \\ 
	 & Platelet distribution width & UK Biobank \citen{sudlow,bycroft,loh} & $\sim 460$ & 17.1 \\ 
	 & RBC distribution width & UK Biobank \citen{sudlow,bycroft,loh} & $\sim 460$ & 19.7 \\ 
	 & RBC count & UK Biobank \citen{sudlow,bycroft,loh} & $\sim 460$ & 17.5 \\ 
	 & White cell count & UK Biobank \citen{sudlow,bycroft,loh} & $\sim 460$ & 20.7 \\ 
	 & Bone mineral density - heel & UK Biobank \citen{sudlow,bycroft,loh} & $\sim 460$ & 29 \\ 
	 & Balding - male*** & UK Biobank \citen{sudlow,bycroft,loh} & $\sim 230$ & 16.1 \\ 
	 & BMI & UK Biobank \citen{sudlow,bycroft,loh} & $\sim 460$ & 27.5 \\ 
	 & Height & UK Biobank \citen{sudlow,bycroft,loh} & $\sim 460$ & 24.7 \\ 
	 & BP - diastolic & UK Biobank \citen{sudlow,bycroft,loh} & $\sim 460$ & 32.3 \\ 
	 & BP - systolic & UK Biobank \citen{sudlow,bycroft,loh} & $\sim 460$ & 28.3 \\ 
	 & College & UK Biobank \citen{sudlow,bycroft,loh} & $\sim 460$ & 19.1 \\ 
	 & Smoking status & UK Biobank \citen{sudlow,bycroft,loh} & $\sim 460$ & 24.9 \\ 
	 & Eczema & UK Biobank \citen{sudlow,bycroft,loh} & $\sim 460$ & 21.8 \\ 
	 & Asthma & UK Biobank \citen{sudlow,bycroft,loh} & $\sim 460$ & 16.8 \\ 
	 & Dermatology & UK Biobank \citen{sudlow,bycroft,loh} & $\sim 460$ & 9.1 \\ 
	 & Myocardial infarction & UK Biobank \citen{sudlow,bycroft,loh} & $\sim 460$ & 18.6 \\ 
	 & High cholesterol & UK Biobank \citen{sudlow,bycroft,loh} & $\sim 460$ & 15.6 \\ 
	 & Hypertension & UK Biobank \citen{sudlow,bycroft,loh} & $\sim 460$ & 36.2 \\ 
	 & Hypothyroidism & UK Biobank \citen{sudlow,bycroft,loh} & $\sim 460$ & 20.1 \\ 
	 & Type 2 Diabetes & UK Biobank \citen{sudlow,bycroft,loh} & $\sim 460$ & 19.5 \\ 
	 & Basal metabolic rate & UK Biobank \citen{sudlow,bycroft,loh} & $\sim 460$ & 23.4 \\ 
	 & FEV1/FVC & UK Biobank \citen{sudlow,bycroft,loh} & $\sim 460$ & 17.7 \\ 
	 & FVC & UK Biobank \citen{sudlow,bycroft,loh} & $\sim 460$ & 18.8 \\ 
	 & Neuroticism & UK Biobank \citen{sudlow,bycroft,loh} & $\sim 460$ & 28.7 \\ 
	 & Morning person & UK Biobank \citen{sudlow,bycroft,loh} & $\sim 460$ & 21.1 \\ 
	 & Age at menarche & UK Biobank \citen{sudlow,bycroft,loh} & $\sim 230$ & 24 \\ 
	 & Age at menopause & UK Biobank \citen{sudlow,bycroft,loh} & $\sim 230$ & 19.1 \\ 
	 & Number children - female & UK Biobank \citen{sudlow,bycroft,loh} & $\sim 230$ & 14.4 \\ 
	 & Number children - male & UK Biobank \citen{sudlow,bycroft,loh} & $\sim 230$ & 15.1 \\
	 \\
	 \caption{52 GWAS datasets included in the analysis. Most UK Biobank summary statistics are publicly available \cite{loh}. All datasets have heritability Z-score $Z_h>7$ and estimated genetic correlation $\hat{\rho}_g<0.9$ with other traits. Summary statistics for $\sim 1,000,000$ HapMap3 SNPs were used, excluding the MHC region. *Total number of samples genotyped by OncoArray; actual sample size is slightly less than 447k. These numbers are excluded from average reported sample size for non-UK Biobank traits. **Actual sample size for UK Biobank analyses is slightly less than 460k (respectively 230k for sex-specific traits), owing to incomplete phenotype data. For most case control traits, effective sample size is substantially less than 460k due to the low fraction of cases. ***The balding phenotype was the ``balding 4" UK Biobank category, corresponding to nearly-complete baldness.}\label{phenotypes-table}
\end{longtable}

}

{\small\centering
\begin{longtable}[H]{c c c c c c c}
\hline 
	Trait~1 & Trait~2 & $p_\text{LCV}$ &  $\hat{\rho}_g$ (std err) & $\gcphat$(std err) & $p_\text{aux}$ &MR ref\\ \hline	
	Triglycerides & Hypertension & $5 \times 10^{-39}$ & 0.25 (0.04) & 0.95 (0.04) & 0.04 \\ 
	BMI & Myocardial infarction & $3 \times 10^{-9}$ & 0.34 (0.09) & 0.94 (0.11) & 0.22& \citen{nordestgaard,lyall}  \\ 
	Triglycerides & Myocardial infarction & $8 \times 10^{-32}$ & 0.30 (0.06) & 0.90 (0.08) & 0.04 & \citen{do}\\ 
	Triglycerides & BP - systolic & $6 \times 10^{-41}$ & 0.13 (0.03) & 0.89 (0.08) & $8 \times 10^{-4}$ \\ 
	HDL & Hypertension & $6 \times 10^{-22}$ & -0.29 (0.06) & 0.87 (0.09) & 0.15 \\ 
	LDL & High cholesterol & $8 \times 10^{-7}$ & 0.77 (0.07) & 0.86 (0.11) & 0.08 \\ 
	Triglycerides & Mean cell volume & $10 \times 10^{-19}$ & -0.20 (0.04) & 0.86 (0.11) & $2 \times 10^{-4}$ \\ 
	Triglycerides & BP - diastolic & $5 \times 10^{-39}$ & 0.11 (0.04) & 0.86 (0.10) & 0.004 \\ 
	Platelet volume & Platelet count & $6 \times 10^{-10}$ & -0.66 (0.03) & 0.84 (0.10) & 0.18 \\ 
	BMI & Hypertension & $2 \times 10^{-16}$ & 0.38 (0.03) & 0.83 (0.11) & 0.06 & \citen{pickrell,lyall} \\ 
	Triglycerides & Platelet dist width & $5 \times 10^{-17}$ & 0.19 (0.04) & 0.81 (0.13) & $7 \times 10^{-5}$ \\ 
	LDL & BMD& $4 \times 10^{-34}$ & -0.12 (0.05) & 0.80 (0.12) & 0.02 \\ 
	BMI & FVC & $4 \times 10^{-13}$ & -0.22 (0.03) & 0.79 (0.17) & 0.001 & \citen{skaaby} \\ 
	Triglycerides & Reticulocyte count & $2 \times 10^{-10}$ & 0.33 (0.05) & 0.79 (0.14) & 0.02 \\ 
	Triglycerides & Eosinophil count & $3 \times 10^{-17}$ & 0.14 (0.05) & 0.75 (0.16) & 0.001 \\ 
	Balding - male & Num children - male & $2 \times 10^{-30}$ & -0.16 (0.05) & 0.75 (0.13) & $2 \times 10^{-4}$ \\ 
	HDL & Platelet dist width & $8 \times 10^{-17}$ & -0.14 (0.04) & 0.75 (0.16) & 0.004 \\ 
	RBC dist width & Type 2 Diabetes & $3 \times 10^{-4}$ & 0.11 (0.03) & 0.73 (0.19) & 0.21 \\ 
	LDL & Myocardial infarction & $2 \times 10^{-31}$ & 0.17 (0.08) & 0.73 (0.13) & $6 \times 10^{-4}$ & \citen{cohen,voight} \\ 
	Platelet dist width & Platelet count & $1 \times 10^{-7}$ & -0.47 (0.04) & 0.73 (0.15) & 0.04 \\ 
	Hypothyroidism & Type 2 Diabetes & $2 \times 10^{-4}$ & 0.22 (0.05) & 0.73 (0.29) & 0.2 \\ 
	HDL & Type 2 Diabetes & $2 \times 10^{-7}$ & -0.40 (0.06) & 0.72 (0.17) & 0.35 \\ 
	Hypothyroidism & Myocardial infarction & $6 \times 10^{-12}$ & 0.26 (0.05) & 0.72 (0.16) & 0.08 \\ 
	High cholesterol & Myocardial infarction & $2 \times 10^{-4}$ & 0.52 (0.12) & 0.71 (0.19) & 0.32 \\ 
	HDL & BP - diastolic & $4 \times 10^{-17}$ & -0.12 (0.06) & 0.70 (0.18) & 0.005 \\ 
	Platelet dist width & Reticulocyte count & $1 \times 10^{-7}$ & 0.13 (0.04) & 0.69 (0.20) & 0.005 \\ 
	LDL & College & $1 \times 10^{-10}$ & -0.13 (0.05) & 0.68 (0.30) & 0.35 \\ 
	Triglycerides & Monocyte count & $1 \times 10^{-4}$ & 0.14 (0.04) & 0.67 (0.21) & 0.09 \\ 
	Type 2 Diabetes & Ulcerative Colitis & $2 \times 10^{-5}$ & -0.14 (0.07) & 0.65 (0.23) & 0.41 \\ 
	BMI & Reticulocyte count & $4 \times 10^{-5}$ & 0.39 (0.03) & 0.64 (0.25) & $10 \times 10^{-4}$ \\ \hline
	HDL & FEV1/FVC & $1 \times 10^{-13}$ & -0.09 (0.04) & 0.56 (0.08) & 0.19 \\ 
	High cholesterol & Neuroticism & $2 \times 10^{-14}$ & 0.09 (0.03) & 0.55 (0.19) & 0.32 \\ 
	Triglycerides & Basal metab rate & $2 \times 10^{-8}$ & 0.08 (0.04) & 0.55 (0.13) & 0.25 \\ 
	Height & BMD& $3 \times 10^{-14}$ & -0.09 (0.04) & 0.50 (0.14) & $2 \times 10^{-8}$ \\ 
	Triglycerides & Height & $3 \times 10^{-14}$ & -0.10 (0.03) & 0.45 (0.09) & 0.15 \\ 
	HbA1C & High cholesterol & $5 \times 10^{-22}$ & 0.25 (0.06) & 0.44 (0.16) & 0.49 \\ 
	Age at menarche & Height & $7 \times 10^{-11}$ & 0.16 (0.04) & 0.43 (0.10) & $2 \times 10^{-5}$ & \citen{pickrell}\\ 
	High cholesterol & Smoking status & $5 \times 10^{-19}$ & 0.13 (0.03) & 0.42 (0.02) & 0.52 \\ 
	Reticulocyte count & Hypertension & $2 \times 10^{-4}$ & 0.27 (0.04) & 0.41 (0.13) & 0.75 \\ 
	BMI & Asthma & $4 \times 10^{-14}$ & 0.21 (0.03) & 0.40 (0.27) & 0.05 & \citen{skaaby} \\ 
	High cholesterol & Monocyte count & $4 \times 10^{-4}$ & 0.09 (0.03) & 0.40 (0.15) & 0.2 \\ 
	Height & Basal metab rate & $10 \times 10^{-9}$ & 0.57 (0.03) & 0.39 (0.07) & 0.006 \\ 
	Eczema & FEV1/FVC & $2 \times 10^{-15}$ & -0.08 (0.03) & 0.36 (0.10) & $2 \times 10^{-5}$ \\ 
	Height & College & $3 \times 10^{-6}$ & 0.17 (0.03) & 0.33 (0.10) & 0.06 & \citen{tyrell}\\ 
	Prostrate cancer & Hypothyroidism & $10 \times 10^{-5}$ & -0.12 (0.05) & 0.30 (0.38) & 0.19 \\ 
	Crohns Disease & LDL & $4 \times 10^{-13}$ & -0.12 (0.06) & 0.29 (0.15) & 0.82 \\ 
	High cholesterol & Type 2 Diabetes & $4 \times 10^{-6}$ & 0.42 (0.05) & 0.24 (0.30) & 0.62 \\ 
	RBC count & Monocyte count & $8 \times 10^{-7}$ & 0.14 (0.05) & 0.24 (0.46) & 0.31 \\ 
	HbA1C & BMI & $7 \times 10^{-17}$ & 0.25 (0.05) & 0.23 (0.35) & 0.77 \\ 
	Basal metab rate & Hypothyroidism & $6 \times 10^{-21}$ & 0.11 (0.04) & 0.21 (0.04) & 0.04 \\ 
	Platelet dist width & Corpuscular hemoglobin & $5 \times 10^{-14}$ & -0.06 (0.02) & 0.15 (0.14) & 0.08 \\ 
	Depressive syndrome & Asthma & $4 \times 10^{-4}$ & 0.21 (0.05) & 0.14 (0.08) & 0.37 \\ 
	BMI & High cholesterol & $2 \times 10^{-6}$ & 0.33 (0.06) & 0.13 (0.12) & 0.25 \\ 
	Age at menopause & Depressive syndrome & $2 \times 10^{-7}$ & -0.27 (0.06) & 0.12 (0.32) & 0.41 \\ 
	White cell count & BMI & $7 \times 10^{-5}$ & 0.24 (0.03) & 0.09 (0.16) & 1 \\ 
	Asthma & Lymphocyte count & $2 \times 10^{-4}$ & 0.09 (0.04) & 0.08 (0.19) & 0.57 \\ 
	Num children - male & Hypothyroidism & $9 \times 10^{-11}$ & 0.18 (0.05) & 0.03 (0.26) & 0.87 \\ 
	College & High cholesterol & $2 \times 10^{-8}$ & -0.23 (0.03) & 0.01 (0.08) & 0.34 \\ 
	RBC dist width & High cholesterol & $4 \times 10^{-4}$ & 0.11 (0.04) & 0.00 (0.17) & 0.35 \\ 
	\caption{Pairs of traits with evidence of partial genetic causality. We restricted to pairs of traits having a nominally significant genetic correlation (two-tailed $p<0.05$; 429 trait pairs) and reported all traits with strong evidence of partial causality (1\% FDR). Trait pairs are ordered so that trait~1 is genetically causal or partially genetically causal for trait~2. We have provided references for each trait pair with existing support in the MR literature that we are aware of. For some trait pairs, there was strong evidence for partial causality but low and noisy $\gcp$ estimates. This phenomenon may occur due to multiple intermediaries, which can cause the estimated mixed fourth moments to have opposite signs. When this occurs, the approximate likelihood function is sometimes bimodal, with no support for any specific value of $\gcp$ (because there is no value of $\gcp$ that produces mixed fourth moments of opposite signs). While this phenomenon appears to occur for several traits with $\gcp$ estimates close to zero, there were no trait pairs with statistically significant evidence that their mixed fourth moments had opposite signs.}\label{all-sig-table}
\end{longtable}
}

\clearpage
\section{LCV model}

The LCV random effects model assumes that the distribution of marginal effect sizes for the two traits can be written as the sum of two independent bivariate distributions (visualized in Figure \ref{diagram-fig}c-e in orange and blue respectively): (1) a {\it shared genetic component} $(q_1\pi,q_2\pi)$ whose values are proportional for both traits; and (2) an {\it even genetic component} $(\gamma_1,\gamma_2)$ whose density is mirror symmetric across both axes. Distribution (1) resembles a line through the origin, and we interpret its effects as being mediated by a latent causal variable ($L$) (Figure \ref{diagram-fig}a); distribution (2) does not contribute to the genetic correlation, and we interpret its effects as direct effects. Informally, the LCV model assumes that any asymmetry in the shared genetic architecture arises from a genetic component that is fully shared between the two traits. 

In detail, the LCV model assumes that there exist scalars $q_1,q_2$, and a distribution $(\pi,\gamma_1,\gamma_2)$ such that
\bal
(\alpha_1,\alpha_2)=(q_1\pi+q_2\pi)+(\gamma_1,\gamma_2),\text{ where } \pi\perp(\gamma_1,\gamma_2)\text{ and } (\gamma_1,\gamma_2){\sim} (-\gamma_1,\gamma_2){\sim}(\gamma_1,-\gamma_2).
\end{align}
Here $\alpha_k$ is the random marginal effect (i.e. the correlation) of a SNP of trait $k$, $\pi$ interpreted as the marginal effect of a SNP on $L$, and $\gamma_k$ is interpreted as the non-mediated effect of a SNP on trait $k$. $\alpha$ and $\pi$ (but not $\gamma$) are normalized to have unit variance, and all random variables have zero mean. (The symbol ``${\sim}$'' means ``has the same distribution as.'') $q_1,q_2$ are the model parameters of primary interest, and we can relate them to the mixed fourth moments, which are observable (equation \eqref{main-eq}). In particular, this implies that the model is identifiable (except when the excess kurtosis $\kappa_\pi=0$). Note that we have avoided assuming a particular parametric distribution.

The LCV model assumptions are strictly weaker than the assumptions made by MR. Like LCV, a formulation of the MR assumptions is that the bivariate distribution of SNP effect sizes can be expressed in terms of two distributions. In particular, it assumes that the effect size distribution is a mixture of (1') a distribution whose values are proportional for both traits (representing all SNPs that affect the exposure Y1) and (2') a distribution with zero values for the exposure Y1 (representing SNPs that only affect the outcome Y2). These two distributions can be compared with distributions (1) and (2) above. Because (1') is identical to (1) and (2') is a special case of (2), the LCV model assumptions are strictly weaker than the MR assumptions (indeed, much weaker). We also note that the MR model is commonly illustrated with a non-genetic confounder affecting both traits. Our latent variable $L$ is a genetic variable, and it is not analogous to the non-genetic confounder. Similar to MR, LCV is unaffected by nongenetic confounders (such a confounder may result in a phenotypic correlation that is unequal to the genetic correlation).

The genetic causality proportion ($\gcp$) is defined as:
\begin{align}
\gcp:=\ff{\log |q_2|-\log |q_1|}{\log |q_2|+\log |q_1|},
\end{align}
which satisfies
\bal
\ff{q_2^2}{q_1^2}=(\rho_g^2)^\gcp,
\end{align}
where the genetic correlation $\rho_g$ is equal to $q_1q_2$. $\gcp$ is positive when trait~1 is partially genetically causal for trait~2. When $\gcp=1$, trait~1 is fully genetically causal for trait~2: $q_1=1$ and the causal effect size is $q_2=\rho_g$ (Figure \ref{diagram-fig}b,e).  The LCV model is broadly related to dimension reduction techniques such as Factor Analysis\cite{child} and Independent Components Analysis \cite{comon}, although it differs in its modeling assumptions as well as its goal (causal inference); our inference strategy (mixed fourth moments) also differs.

Under the LCV model assumptions, we derive equation \eqref{main-eq} as follows:
\bal\label{main-eq-derivation}
E(\alpha_1^3\alpha_2)=&E((\gamma_1+q_1\pi)^3(\gamma_2+q_2\pi))\nonumber\\
=&q_1^3q_2E(\pi^4)+3q_1q_2E(\pi^2\gamma_1^2)\nonumber\\
=&q_1^3q_2E(\pi^4)+3q_1q_2E(\pi^2)E(\gamma_1^2)\nonumber\\
=&q_1^3q_2E(\pi^4)+3q_1q_2(1)(1-q_1^2)\nonumber\\
=&q_1^3q_2(E(\pi^4)-3)+3q_1q_2.
\end{align}
In the second line, we used the independence assumption to discard cross-terms of the form $\gamma_p\pi^3$ and $\gamma_1^3\pi$, and we used the symmetry assumption to discard terms of the form $\gamma_1\gamma_2^3$. In the third and fourth lines, we used the independence assumption, which implies that  $E(\gamma_1^2\pi^2)=E(\gamma_1^2)E(\pi^2)=E(\gamma_1^2)=1-q_1^2$. The factor $E(\pi^4)-3$ is the excess kurtosis of $\pi$, which is zero when $\pi$ follows a Gaussian distribution; in order for equation \eqref{main-eq} to be useful for inference, $E(\pi^4)-3$ must be nonzero, and in order for the model to be identifiable, $\pi$ must be non-Gaussian.

\section{Estimation under the LCV model}
In order to estimate the gcp and to test for partial causality, we utilize six steps. First, we use LD score regression \cite{bulik2} to estimate the heritability of each trait; these estimates are used to normalize the summary statistics. Second, we apply cross-trait LD score regression \cite{bulik1} to estimate the genetic correlation; the intercept in this regression is also used to correct for possible sample overlap when estimating the mixed fourth moments. Third, we estimate the mixed fourth moments $E(\alpha_1\alpha_2^3)$ and $E(\alpha_1^3\alpha_2)$ of the bivariate effect size distribution. Fourth, we compute test statistics for each possible value of the $\gcp$, based on the estimated genetic correlation and on the estimated mixed fourth moments. Fifth, we jackknife on these test statistics to estimate their standard errors, similar to ref. \citen{bulik2}, obtaining a likelihood function for the gcp. Sixth, we obtain posterior means and standard errors for the gcp using this likelihood function and a uniform prior distribution. These steps are detailed below.

First, we apply LD score regression to normalize the test statistics. Under the LCV model, the marginal effect sizes for each trait, $\alpha_1$ and $\alpha_2$, have unit variance. We use a slightly modified version of LD score regression \cite{bulik2}, with LD scores computed from UK10K data \cite{uk10k}. In particular, we run LD score regression using a slightly different weighting scheme, matching the weighting scheme in our mixed fourth moment estimators; the weight of SNP $i$ was:
\bal\label{weights}
w_i:=\max(1,1/\ell_i^\text{HapMap}),
\end{align}
where $\ell_i^\text{HapMap}$ was the estimated LD score between SNP $i$ and other HapMap3 SNPs (this is approximately the set of SNPs that were used in the regression). This weighting scheme is motivated by the fact that SNPs with high LD to other regression SNPs will be over-counted in the regression (see ref. \citen{bulik2}).
Similar to ref. \citen{bulik1}, we improve power by excluding large-effect variants when computing the LD score intercept; for this study, we chose to exclude variants with $\cs$ statistic $30\times$ the mean (but these variants are not excluded when computing $\bar{\cs}$). Then, we divide the summary statistics by $s=\sqrt{\bar{\cs}-\hat{\sigma}^2_\epsilon}$, where $\bar{\cs}$ is the weighted mean $\cs$ statistic and $\hat{\sigma}^2_\epsilon$ is the LD score intercept. (We also divide the LD score intercept by $s^2$.) We assess the significance of the heritability by performing a block jackknife on $s$, defining the significance $Z_h$ as $s$ divided by its estimated standard error. 

Second, to estimate the genetic correlation, we apply cross-trait LD score regression \cite{bulik1}. Similar to above, we use a slightly modified weighting scheme (equation \eqref{weights}), and we exclude large-effect variants when computing the cross-trait LD score intercept. We assess the significance of the genetic correlation using a block jackknife. 

Third, we estimate the mixed fourth moments $E(\alpha_1\alpha_2^3)$ using the following estimation equation:
\bal
E(a_1a_2^3|\alpha_1,\alpha_2)=&\aa_1\aa_2^3+E(\epsilon_1\epsilon_2^3)+3E(\aa_1\aa_2\epsilon_2^2)+E(\aa_2^2\epsilon_1\epsilon_2)\nonumber\\
=&\aa_1\aa_2^3+3E(\epsilon_1\epsilon_2)E(\epsilon_2^2)+3\aa_1\aa_2E(\epsilon_2^2)+\alpha_2^2E(\epsilon_1\epsilon_2)\nonumber\\
=&\aa_1\aa_2^3+3\hat{\sigma}_{\epsilon_1\epsilon_2}\hat{\sigma}^2_{\epsilon_2}+3\aa_1\aa_2\hat{\sigma}^2_{\epsilon_2}+\alpha_2^2\hat{\sigma}_{\epsilon_1\epsilon_2},
\end{align}
where $E(\epsilon_k^2)$ is the LD score regression intercept for trait $k$ and $\hat{\sigma}_{\epsilon_1\epsilon_2}$ is the cross-trait LD score regression intercept. For simulations with no LD, we use $E(\epsilon_k^2)= 1/sN_k$ and $E(\epsilon_1\epsilon_2)=0$ instead of estimating these values.

Fourth, we define a collection of statistics $S(x)$ for $x\in X=\{-1,-.01,-.02,...,1\}$ (corresponding to possible values of $\gcp$):
\bal\label{S-eq}
S(x):=\ff{A(x)-B(x)}{\max(1/|\hat{\rho}_g|,\sqrt{A(x)^2+B(x)^2})}\quad A(X)=|\rho_g|^{-x}\hat{k}_1,\quad B(x)=|\rho_g|^{x}\hat{k}_2,
\end{align}
The motivation for utilizing the normalization by $\sqrt{A(x)^2+B(x)^2}$ is that the magnitude of $A(x)$ and $B(x)$ tend to be highly correlated, leading to increased standard errors if we only use the numerator of $S$. However, the denominator tends to zero when the genetic correlation is zero, leading to instability in the test statistic and false positives. The use of the threshold leads to conservative, rather than inflated, standard errors when the genetic correlation is zero or nearly zero. We recommend only analyzing trait pairs with a  significant genetic correlation, and this threshold usually has no effect on the results. It is also inadvisable to analyze trait pairs whose genetic correlation is non-significant because for positive LCV results, the genetic correlation provides critical information about the causal effect size and direction. 

Fifth, we estimate the variance of $S(x)$ using a block jackknife with $k=100$ blocks of contiguous SNPs, resulting in minimal non-independence between blocks. Blocks are chosen to include the same number of SNPs, and the jackknife standard error is 
\bal
\hat{\sigma}_{S(x)}=\sqrt{101\sum_{j=1}^{100} (S_j(x)-\bar{S}(x))^2}
\end{align}
where $S_j(x)$ is the test statistic computed on blocks $1,...,j-1,j+1,...100$ and $\bar{S}(x)$ is the mean of the jackknife estimates. We compute an approximate likelihood, $L(S|\gcp=x)$, by assuming (1) that $L(S|\gcp=x)=L(S(x)|\gcp=x)$ and (2) that if $\gcp=x$ then $S(x)/\hat{\sigma}_{S(x)}$ follows a T distribution with 98 degrees of freedom. 

Sixth, we impose a uniform prior on $\gcp$, enabling us to obtain a posterior mean estimate of the $\gcp$:
\bal
\gcphat:=\ff{1}{|X|}\sum_{x\in X}xL(x)
\end{align}
The estimated standard error is:
\bal
\hat{\text{se}}:=\sqrt{\ff{1}{|X|}\sum_{x\in X}(x-\gcphat)^2L(x)}.
\end{align}
In order to compute p-values, we apply a T-test to the statistic $S(0)$.

\section{LCV model violations}

In this section, we define partial genetic causality without making LCV (or other) model assumptions and characterize the type of LCV model violation that causes LCV to produce false positives and bias. There are two classes of LCV model violations: {\it independence violations} and {\it proportionality violations}. Roughly, independence violations involve a violation of the independence assumption between mediated effects ($\pi$) and direct effects ($\gamma$) while still satisfying a key proportionality condition related to the mixed fourth moments; as a result, independence violations are not expected to cause LCV to produce false positives. Proportionality violations, on the other hand, violate this proportionality condition and are potentially more problematic. In order to make this characterization, it is necessary to define partial genetic causality in a more general setting, without assuming the LCV model. Partial genetic causality is defined in terms of the {\it correlated genetic component} of the bivariate SNP effect size distribution, which generalizes the shared genetic component modeled by LCV; unlike the shared genetic component, the correlated genetic component does not have proportional effects on both traits (but merely correlated effects). 

\subsection{Definition of partial genetic causality without LCV model assumptions} 
Let $A=(\alpha_1,\alpha_2)$ be the bivariate distribution of marginal effect sizes, normalized to have zero mean and unit variance. First, we define an {\it even genetic component} of $A$ as a distribution $T=(t_1,t_2)$ that is independent of its complement $A-T$ and that satisfies a mirror symmetry condition:
\bal
(t_1,t_2)\sim(-t_1,t_2)\sim(t_1,-t_2).
\end{align}
Equivalently, the density function of $T$ is an even function of both variables. Note that an even genetic component does not contribute to the genetic correlation. In order to define the ``correlated genetic component,'' we would like to define a maximal even component, i.e. an even component that explains the largest possible amount of heritability for both traits. However, if $A$ follows a Gaussian distribution, then there is no maximal even component: instead, the even genetic component that maximizes the proportion of trait~1 heritability explained fails to maximize the proportion of trait~2 heritability explained. This fact is related to the observation that the LCV model is non-identifiable when the effect size distribution for $L$ follows a Gaussian distribution, and {\it only} when it follows a Gaussian distribution\cite{oconnor}. Generalizing this result, we conjecture that there exists an even component that is maximal {up to a Gaussian term}. More precisely, there exists a maximal even component $T^*=(t_1^*,t_2^*)$ such that for any even component $T=(t_1,t_2)$, there exists a (possibly degenerate) Gaussian random variable $Z=(z_1^*,z_2^*)$ independent of $T^*$ such that $T^*+Z$ is an even component and $E((t_1^*+z_1)^2)\geq E(t_1^2)$ and $E((t_2^*+z_2)^2)\geq E(t_2^2)$.

We define the {\it correlated genetic component} $S=(s_1,s_2)$ as the complement of the maximal even component and the Gaussian term.  Trait~1 is defined as {\it partially genetically causal} for trait~2 if $E(s_1^2)>E(s_2^2)$, and vice versa. We may also define the genetic causality proportion using main text equation (1), substituting $E(s_k^2)$ for $q_k^2$. However, the interpretation of the gcp is not as clear in this more general setting. Note that the correlated genetic component may be identically 0, for example if $A$ is bivariate Gaussian or if $A$ itself is an even component; in both cases, there is no partial causality, and the genetic causality proportion is undefined. In practice, if the correlated genetic component is 0 or nearly 0, LCV will produce null p-values and low, noisy gcp estimates.

\subsection{Independence violations and proportionality violations}
The LCV model assumption is equivalent to the statement that the correlated genetic component is equal to a shared genetic component: $S=(q_1\pi, q_2\pi)$, for some random variable $\pi$ and fixed parameters $q_1,q_2$ such that $\rho_g=q_1q_2$. This assumption enables an inference approach based on mixed fourth moments because it implies that the mixed fourth moments of the correlated component are proportional to the respective variances:
\bal\label{proportionality-condition}
E(s_1s_2s_k^2)\propto E(s_k^2),
\end{align}
where under the LCV model, the proportionality constant is $q_1q_2E(\pi^4)$. One form of LCV model violation arises when $S$ is not a shared genetic component, but \eqref{proportionality-condition} still holds.
Intuitively, this type of violation arises as a result of non-independence between mediated effects ($\pi$) and direct effects ($\gamma$), causing ``noise'' from the direct effects to be incorporated into the correlated component, and  we call such violations {\it independence violations}; genetic architectures that violate the proportionality condition we call {\it proportionality violations}. In the presence of an independence violation, we obtain the following moment condition, generalizing main text equation (2):
\bal
E(\alpha_1\alpha_2\alpha_k^2)=cE(s_k^2)+3\rho_g
\end{align}
where $c$ is a proportionality constant. In particular, if $E(s_1^2)=E(s_2^2)$ (no partial causality), then $E(\alpha_1\alpha_2^3)=E(\alpha_2\alpha_1^3)$, and LCV is expected to produce well-calibrated p-values. Conversely, under a proportionality violation, LCV is expected to produce inflated p-values under the null.

 However, the interpretation of the gcp is not as clear in this more general setting; in particular, a gcp of 1 implies that every SNP affecting trait~1 also affects trait~2, but not proportionally. Note that the correlated genetic component may be identically 0, for example if $A$ is bivariate Gaussian or if $A$ itself is an even genetic component; in both cases, there is no partial causality, and the genetic causality proportion is undefined. In practice, if the correlated genetic component is 0 or nearly 0, LCV will produce null p-values and low, noisy gcp estimates.

\section{Extended Simulations}
\subsection{Existing Mendelian Randomization methods}

\noindent {\bf Two-sample MR.} As described in ref. \citen{burgess0}, we ascertained significant SNPs ($p<5\times 10^{-8}$, $\cs$ test) for the exposure and performed an unweighted regression, with intercept fixed at zero, of the estimated effect sizes on the outcome with the estimated effect sizes on the exposure (in practice, a MAF-weighted and LD-adjusted regression is often used; in our simulations, all SNPs had equal MAF, and there was no LD). To assess the significance of the regression coefficient, we estimated the standard error as $\text{se}=\sqrt{\ff{\ff{1}{K}\sum_{k=1}^K \bar{\beta}_{k2}^2}{\sum_{k=1}^K \hat{\beta}_{k1}^2}}$, where $ \bar{\beta}_{k2}$ is the $k^\text{th}$ residual, $N_2$ is the sample size in the outcome cohort, and $K$ is the number of significant SNPs. This estimate of the standard error allows the residuals to be overdispersed compared with the error that is expected from the GWAS sample size. To obtain p values, we applied a two-tailed $t$-test to the regression coefficient divided by its standard error, with $K-1$ degrees of freedom.

\vspace{.25cm}\noindent {\bf MR-Egger.} As described in ref. \citen{bowden}, we ascertained significant SNPs for the exposure and coded them so that the alternative allele had a positive estimated effect on the exposure. We performed an unweighted regression with a fitted intercept of the estimated effect sizes on the outcome on the estimated effect sizes on the exposure. We assessed the significance of the regression using  the same procedure as for two-sample MR, except that the $t$-test used $K-2$ rather than $K-1$ degrees of freedom.

\vspace{.25cm}\noindent {\bf Bidirectional MR.} We implemented bidirectional mendelian randomization in a manner similar to ref. \citen{pickrell}. Significant SNPs were ascertained for each trait. If the same SNP was significant for both traits, then it was assigned only to the trait where it ranked higher (if a SNP ranked equally high for both traits, it was excluded from both SNP sets). The Spearman correlations $r_1,\, r_2$ between the $z$ scores for each trait was computed on each set of SNPs, and we applied a $\cs_1$ test to
\bal
\cs=\ff{1}{\ff{1}{K_1-3}+\ff{1}{K_2-3}}(\text{atanh}(r_1)-\text{atanh}(r_2))^2,
\end{align}
where $K_j$ is the number of significant SNPs for trait $j$. In ref. \citen{pickrell}, the statistics $\text{atanh}(r_j)$ were also used, but a relative likelihood comparing several different models was reported instead of a p-value. We chose to report p-values for Bidirectional MR in order to allow a direct comparison with other methods.

\vspace{.25cm}\noindent {\bf Application of MR to real data.} For our applications of MR and related methods to real data, we selected genetic instruments using a greedy pruning procedure. We ranked all genome-wide significant SNPs for the exposure ($p<5\times 10^{-8}$) by $\cs$ statistic. Iteratively, we removed all SNPs within 1cM of the first SNP in the list, obtaining a set of independent lead SNPs separated by at least 1cM. We confirmed using an LD reference panel that our 1cM window was sufficient to minimize LD among the set of retained SNPs. We applied each MR method as described above; in particular, we performed unweighted regressions for MR and MR-Egger.

\begin{figure}[h!]
\includegraphics[width=\textwidth]{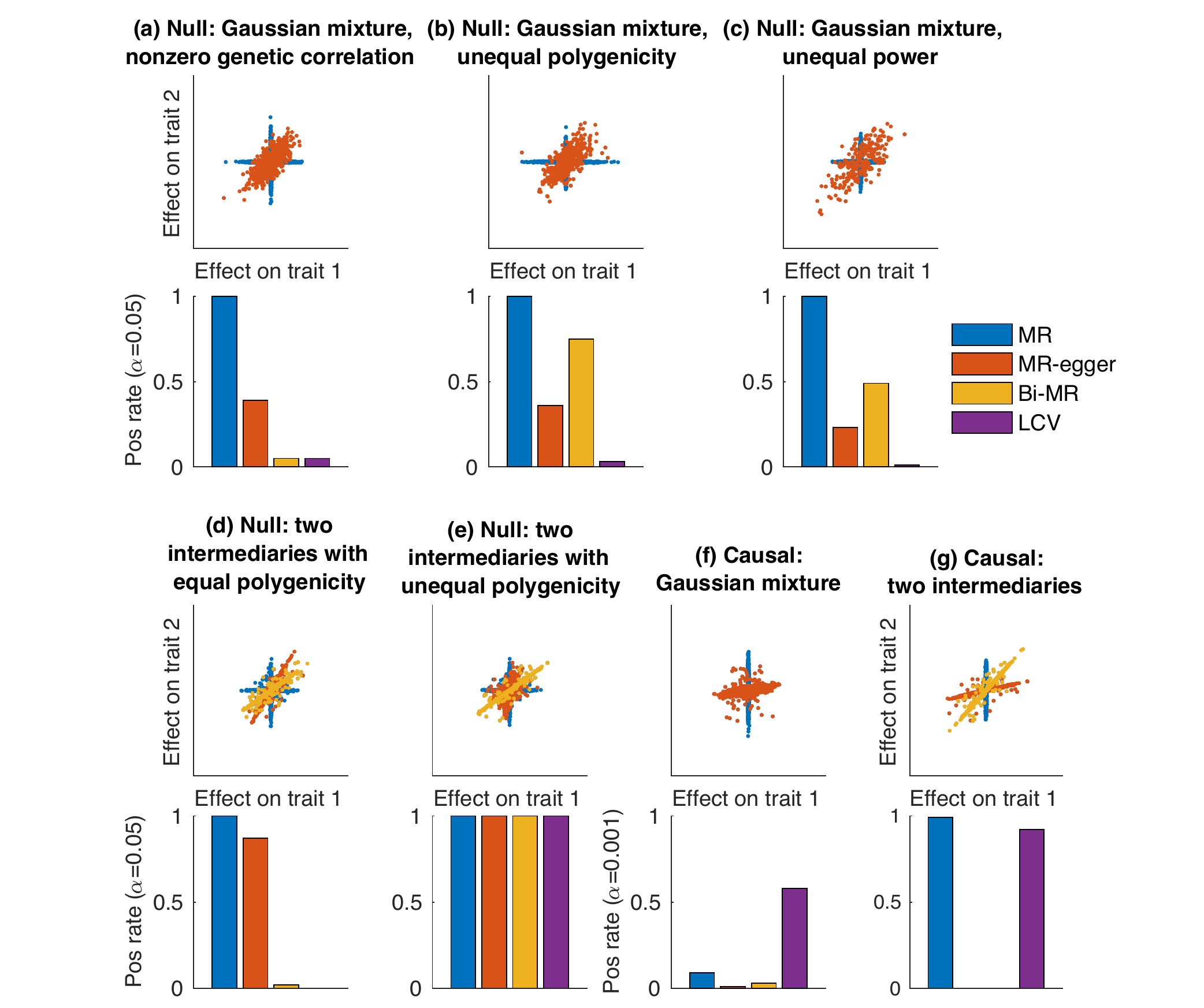}
\caption{Null and causal simulations with no LD and LCV model violations. We report the positive rate ($\alpha=0.05$ for null simulations, $\alpha=0.001$ for causal simulations) for two-sample MR, MR-Egger, Bidirectional MR and LCV. Panels (a)-(c) correspond to Gaussian mixture model extensions of the models in Figure \ref{MR-null-fig}b-d. Panels (f) and (g) correspond to causal analogues of the models in panels (a) and (d), respectively. We also display scatterplots illustrating the bivariate distribution of true SNP effect sizes on the two traits. 
(a) Null simulation with nonzero SNP effects drawn from a mixture of Gaussian distributions; one mixture component has correlated effects on each trait.
(b) Null simulation with SNP effects drawn from a mixture of Gaussian distributions, and differential polygenicity between the two traits.
(c) Null simulation with SNP effects drawn from a mixture of Gaussian distributions, and unequal power between the two traits.
(d) Null simulation with two intermediaries having different effects on each trait. 
(e) Null simulation with two intermediaries having different effects on each trait and unequal polygenicity for the two intermediaries. 
(f) Causal simulation with SNP effects drawn from a mixture of Gaussian distributions; all SNPs affecting trait~1 also affect trait~2, but the relative effect sizes were noisy.
(g) Causal simulation with an additional genetic confounder (i.e. a second intermediary) mediating part of the genetic correlation.  
Results for each panel are based on 1000 simulations. }\label{multi-intermediary-fig}
\end{figure}
\begin{figure}[h!]\centering
\includegraphics[width=\textwidth]{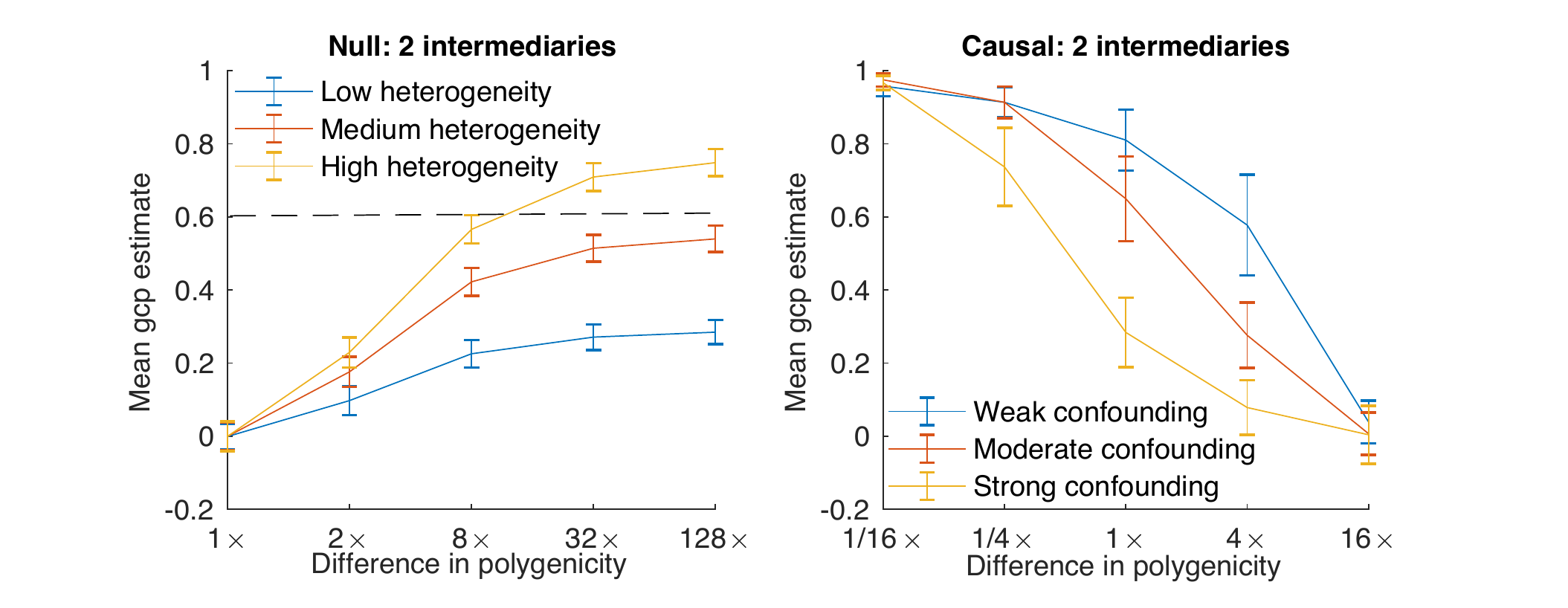}
\caption{Mean gcp estimates in simulations with LCV model violations (error bars show standard deviations; based on 1000 simulations). (a) Null simulation with two intermediaries having possibly unequal polygenicity. The two intermediaries had either a slightly, moderately, or highly heterogenous effect on the two traits; that is, when heterogeneity was high, intermediary 1 had a much larger effect on trait 1 while intermediary 2 had a much larger effect on trait 2. Then, we specified a certain difference in polygenicity between the two traits (measured by the proportion of causal SNPs). (b) Casual simulation with an additional latent confounder. The latent confounder explained either a low, medium or high proportion of the genetic correlation. We varied the polygenicity of the confounder and of the causal trait, such that a 16x difference in polygenicity indicates that 16x more SNPs were causal for the causal trait than for the genetic confounder. }\label{gcp-misspec-fig}
\end{figure}

\subsection{Simulations involving LCV model violations}

In order to investigate potential limitations of our approach, we performed null and causal simulations under genetic architectures that violate LCV model assumptions.  There are two classes of LCV model violations: {\it independence violations} and {\it proportionality violations} (see LCV model violations). Roughly, independence violations involve a violation of the independence assumption between (1) mediated effects ($\pi$) and (2) direct effects ($\gamma$) while still satisfying a key proportionality condition related to the mixed fourth moments; as a result, independence violations are not expected to cause LCV to produce false positives. Proportionality violations, on the other hand, violate this proportionality condition and are potentially more problematic.
A representative example of an independence violation is a bivariate Gaussian mixture model where one of the mixture components generates imperfectly correlated effect sizes on the two traits. These SNPs underlying this mixture component can be viewed as having both an effect on $L$ and also a residual effect on the two traits directly, in violation of the independence assumption. First, we performed null simulations under a Gaussian mixture model with a nonzero genetic correlation. These simulations were similar to the simulations reported in Figure 2b, except that the correlated SNP effect sizes (1\% of SNPs) were drawn from a bivariate normal distribution with correlation 0.5 (explaining 20\% of heritability for each trait; in Figure 2b, these effects were perfectly correlated). Similar to Figure 2b, LCV and bidirectional MR produced p-values that were well-calibrated, while MR and MR-Egger produced inflated p-values (Figure~\ref{multi-intermediary-fig}a). Second, similar to Figure 2c, we included differential polygenicity between the two traits, finding that differential polygenicity caused all existing methods including bidirectional MR, but not LCV, to produce false positives (Figure~\ref{multi-intermediary-fig}b). Third, similar to Figure 2d, we included differential power between the two traits, again finding that LCV produced well-calibrated p-values while existing methods produced false positives (Figure~\ref{multi-intermediary-fig}c). 

A representative example of a proportionality violation is a model in which two intermediaries $L_1$ and $L_2$ have different effect sizes on the two traits, and $L_1$ and $L_2$ also have unequal polygenicity. First, for comparison purposes, we considered a model with two intermediaries with equal polygenicity; 2\% of SNPs were causal for each intermediary, and 4\% of SNPs were causal for each trait exclusively. Because this model implies only an independence violation, we expected that LCV would not produce false positives. Indeed, LCV produced well-calibrated p-values (Figure~\ref{multi-intermediary-fig}d). Similar to Figure 2b and Figure~\ref{multi-intermediary-fig}a, Bidirectional MR also produced well-calibrated p-values, while MR and MR-Egger produced false positives. Second, we shifted the polygenicity of the two intermediaries in opposite directions: 1\% of SNPs were causal for $L_1$ and 8\% of SNPs were causal for $L_2$, resulting in a proportionality violation. We expected that LCV would produce false positives, as the intermediary with lower polygenicity would disproportionately affect the mixed fourth moments. Indeed, LCV (as well as other methods) produced false positives, indicating that proportionality violations cause LCV to produce false positives (Figure~\ref{multi-intermediary-fig}e). We investigated the gcp estimates produced by LCV in these simulations, finding that LCV produced low gcp estimates ($\gcphat\approx 0.5$; Figure~\ref{gcp-misspec-fig}a). We varied the difference in polygenicity as well as the difference in the relative effect sizes of the two intermediaries, finding that extreme parameter settings (e.g., a $32\times$ difference in polygenicity in conjunction with a $25\times$ difference in the relative effect sizes of $L_1$ and $L_2$) were required to cause LCV to produce high gcp estimates ($\gcp>0.6$; Figure~\ref{gcp-misspec-fig}a). Thus, proportionality violations of LCV model assumptions can cause LCV (and other methods) to produce false positives, but genetic causality remains the most parsimonious explanation for high gcp estimates.

Finally, we performed (fully) causal simulations under LCV model violations.  First, we simulated an independence violation by specifying a Gaussian mixture model where every SNP affecting trait~1 also affected trait~2, but the relative effect sizes were noisy (Figure~\ref{multi-intermediary-fig}f). Sample size and polygenicity were similar to Figure 3a ($4\times$ lower sample size than Figure~\ref{multi-intermediary-fig}a). As expected, LCV had lower power to detect a causal effect than in Figure 3a, although it still had moderately high power. Second, we simulated a proportionality violation by specifying both a causal effect (corresponding to $L_1$) and an additional genetic confounder (corresponding to $L_2$) (Figure~\ref{multi-intermediary-fig}g). LCV had lower power to detect a causal effect than in Figure~\label{MR-power-fig}a, although it still had high power.

In summary, we determined in null simulations that independence violations do not cause LCV to produce false positives; in addition, these simulations recapitulated the limitations of existing methods that we observed in simulations under the LCV model. Proportionality violations caused LCV (as well as existing methods) to produce false positives; however, extreme values of the simulation parameters were required in order for LCV to produce high gcp estimates. In causal simulations, we determined that both independence and proportionality violations lead to reduced power for LCV (and other methods), as well as downwardly biased gcp estimates for LCV.

{\small\centering
\begin{longtable}{c c c c c c c c c c c}
	   & 	& $\rho$ & $p<.05$ & $p<.001$ & Mean $\cs$ & Mean $\gcphat$ & $\gcphat$ std dev & RMS $\hat{\sigma}$ & $Z_h$\\ \hline
	a & Zero genetic correlation & 0 & 0 & 0 & 0.32 & -0.00 & 0.11 & 0.55 & 8 \\ 
	b & Low genetic correlation & 0.1 & 0.009 & 0 & 0.58 & 0.00 & 0.14 & 0.29 & 8.5 \\ 
	c & Default parameter values & 0.2 & 0.058 & 0.003 & 1.09 & -0.00 & 0.07 & 0.08 & 8.6 \\ 
	d & High genetic correlation & 0.4 & 0.067 & 0.004 & 1.2 & -0.00 & 0.1 & 0.11 & 8 \\ 
	e & Very high genetic correlation & 0.8 & 0.058 & 0.002 & 1.13 & -0.00 & 0.21 & 0.24 & 5.8 \\ 
	f & Uncorrelated pleiotropic effects & 0.2 & 0.054 & 0.001 & 1.06 & -0.00 & 0.08 & 0.09 & 8.7 \\ 
	g & Differential polygenicity & 0.2 & 0.062 & 0.002 & 1.1 & -0.01 & 0.08 & 0.08 & 10 \\ 
	h & Very different polygenicity & 0.2 & 0.067 & 0.004 & 1.19 & -0.01 & 0.1 & 0.1 & 11.2 \\ 
	i & Low $N_1$ & 0.2 & 0.063 & 0.004 & 1.14 & 0.01 & 0.12 & 0.13 & 5 \\ 
	j & Very low $N_1$ & 0.2 & 0.228 & 0.132 & 11.2 & 0.11 & 0.35 & 0.33 & 1.4 \\ 
	k & Different heritability & 0.2 & 0.061 & 0.005 & 1.7 & 0.00 & 0.09 & 0.1 & 6.5 \\ 
	l & High phenotypic correlation & 0.2 & 0.057 & 0.002 & 1.12 & 0.00 & 0.07 & 0.08 & 8.7 \\ 
	m & Zero phenotypic correlation & 0.2 & 0.057 & 0.005 & 1.1 & 0.00 & 0.07 & 0.08 & 8.6 \\ 
	n & Uncorrelated pleiotropic effects & 0 & 0.001 & 0 & 0.3 & 0.00 & 0.14 & 0.52 & 8 \\ 
	o & Differential polygenicity & 0 & 0 & 0 & 0.31 & -0.02 & 0.12 & 0.55 & 9.8 \\ 
	p & Very different polygenicity & 0 & 0.001 & 0 & 0.31 & -0.05 & 0.14 & 0.52 & 11.4 \\ 
	q & Low $N_1$ & 0 & 0.001 & 0.001 & 0.31 & 0.00 & 0.14 & 0.52 & 5 \\   
	r & Very low $N_1$ & 0 & 0.272 & 0.216 & 46.4 & 0.27 & 0.32 & 0.39 & 1.4 \\ 
	s & Different heritability & 0 & 0 & 0 & 0.28 & -0.00 & 0.11 & 0.55 & 6.3 \\ \hline
	t & Causal & 0.2 & 0.965 & 0.94 & 258 & 0.76 & 0.12 & 0.16 & 8.6 \\ 
	u & Partially causal & 0.2 & 0.706 & 0.347 & 12.9 & 0.56 & 0.15 & 0.24 & 10 \\ 
	v & Low $N_1$ & 0.2 & 0.852 & 0.768 & 66 & 0.65 & 0.17 & 0.2 & 5.1 \\ 
	w & Very low $N_1$ & 0.2 & 0.452 & 0.378 & 102 & 0.39 & 0.35 & 0.35 & 1.4 \\ 
	x & Low $N_2$ & 0.2 & 0.843 & 0.714 & 40.8 & 0.60 & 0.18 & 0.21 & 8.7 \\ 
	y & Weak causal effect & 0.1 & 0.422 & 0.104 & 6.36 & 0.49 & 0.18 & 0.32 & 8.7 \\ 
	z & $Y_1$ less polygenic & 0.2 & 0.997 & 0.996 & 7331 & 0.90 & 0.08 & 0.07 & 3.6 \\ 
	aa & $Y_1$ more polygenic & 0.2 & 0.155 & 0.004 & 2.39 & 0.28 & 0.2 & 0.47 & 13.3 \\ 
	bb & $Y_1$ infinitessimal & 0.2 & 0.012 & 0 & 0.7 & 0.07 & 0.2 & 0.5 & 14.2 \\
	\\ 
	
\caption{Null and non-null simulations with LD. Proportion of simulations (out of $n=5000$) with LCV p-value for partial causality less than $0.05$ and less than $0.001$; mean $\cs$ statistic;  mean $\gcphat$ (in each case, standard error is less than 0.01); empirical standard deviation of $\gcphat$; root mean squared estimated standard error; mean heritability Z-score for trait~1. Simulations a-s are null ($\gcp=0$), and simulations t-bb are non-null. 
(a-e) Different values of the genetic correlation ($\rho$). When the genetic correlation is zero or near-zero, we observe conservative p-values and overestimates of the $\gcphat$ standard error.
(f) Uncorrelated pleiotropic effects: 0.3\% of SNPs affect both traits with independent effect sizes. 
(g-h) Differential or very different polygenicity: 0.2\% and 0.8\% of SNPs, or 0.1\% and 1.6\% of SNPs respectively, have direct effects on each trait.
(i-j) Low or very low sample size for trait~1: either $N_1=20$k or $N_1=4$k respectively, and $N_2=100$k. 
(k) Different heritability: $h^2_1=0.1$ and $h^2_2=0.5$.
(l) High phenotypic correlation of 0.4, compared with $\rho=0.2$.
(m) Zero phenotypic correlation.
(n) Uncorrelated pleiotropic effects: 0.3\% of SNPs affect both traits with independent effect sizes. 
(o-p) Differential or very different polygenicity: 0.2\% and 0.8\% of SNPs, or 0.1\% and 1.6\% of SNPs respectively, have direct effects on each trait. 
(q-r) Low or very low sample size for trait~1: either $N_1=20$k or $N_1=4$k respectively, and $N_2=100$k. 
(s) Different heritability: $h^2_1=0.1$ and $h^2_2=0.5$.
(t) Causal.
(u) Partially causal ($\gcp=0.5$).
 (v-w) Causal, with low or very low sample size for the causal trait ($N_1=20$k or $N_1=4$k, and $N_2=100$k). 
 (x) Causal, with low sample size in the downstream trait ($N_2=20$k, $N_1=100$k).
 (y) Weak causal effect (0.1 rather than 0.25). 
 (z-bb) Varying polygenicity for the causal trait: instead of $0.5\%$ of SNPs causal, either 0.05\%, 5\%, or 100\% of SNPs causal for z-bb respectively.
 }\label{ld-supp-sims-table}
\end{longtable}
}
\begin{table}[h!]\small
\begin{tabular}{l  c l  c l  c l  c l  c l  c l  c l  c l  c l  c l  c l  c l  c l  c }
\hline
	   & 	& $\rho$ & $p<.05$ & $p<.001$ & Mean $\cs$ & Mean $\gcphat$ & $\gcphat$ std dev & RMS $\hat{\sigma}$ & $Z_h$\\ \hline
	a & Default parameter values & 0.2 & 0.034 & 0 & 0.9 & 0.00 & 0.05 & 0.06 & 16.7 \\ 
	b & Zero genetic correlation & 0 & 0.001 & 0 & 0.31 & -0.00 & 0.11 & 0.55 & 15.3 \\ 
	c & Very high genetic correlation & 0.8 & 0.033 & 0.002 & 0.94 & -0.00 & 0.11 & 0.43 & 10.9 \\ 
	d & Uncorrelated pleiotropic effects & 0.2 & 0.032 & 0.000 & 0.87 & -0.00 & 0.06 & 0.09 & 16.6 \\ 
	e & Differential polygenicity & 0.2 & 0.034 & 0.002 & 0.86 & -0.00 & 0.05 & 0.07 & 19.3 \\ 
	f & Low $N_1$ & 0.2 & 0.042 & 0.002 & 0.9 & 0.00 & 0.1 & 0.12 & 8.7 \\ 
	g & Very low $N_1$ & 0.2 & 0.254 & 0.16 & 19.96 & 0.08 & 0.35 & 0.32 & 2.2 \\  \hline
	h & Causal & 0.2 & 0.968 & 0.943 & 257.17 & 0.76 & 0.11 & 0.16 & 16.5 \\ 
	i & Partially causal & 0.2 & 0.765 & 0.369 & 12.54 & 0.57 & 0.15 & 0.23 & 19.2 \\  
	\end{tabular}
\caption{Simulations with LD using constrained-intercept LD score regression to estimate the heritability. This heritability estimation method is less noisy than variable-intercept LD score regression but can produce biased estimates on real data due to population stratification and cryptic relatedness\cite{bulik2}. Proportion of simulations (out of $n=2000$) with p-value for partial causality less than $.05$ and less than $.001$; mean $\cs$ statistic for partial causality; mean $\gcphat$; standard deviation of $\gcp$ estimates; root-mean squared estimated standard error. Simulations a-f are null ($\gcp=0$), and simulations g-h are non-null. (a) Realistic simulation parameters (see Methods). 
(b) Genetic correlation $\rho=0$. 
(c) Genetic correlation $\rho=0.75$. 
(d) Uncorrelated pleiotropic effects in addition to a genetic correlation: 50\% of SNPs with direct (non-mediated) effects on each trait are shared between the two traits. 
(e) Differential polygenicity: 0.2\% and 0.05\% of SNPs have direct effects on each trait. 
(f) Different sample size: $N_1=1000$k and $N_2=500$k. 
(g) Different sample size: $N_1=20$k and $N_2=500$k. 
(h) Full genetic causality: $\gcp=1$, with causal effect equal to the genetic correlation (0.25). 
(i) Partial genetic causality: $\gcp=0.5$.
}\label{fixedint-table}
\end{table}

\begin{figure}[h]\centering
\includegraphics[width=\textwidth]{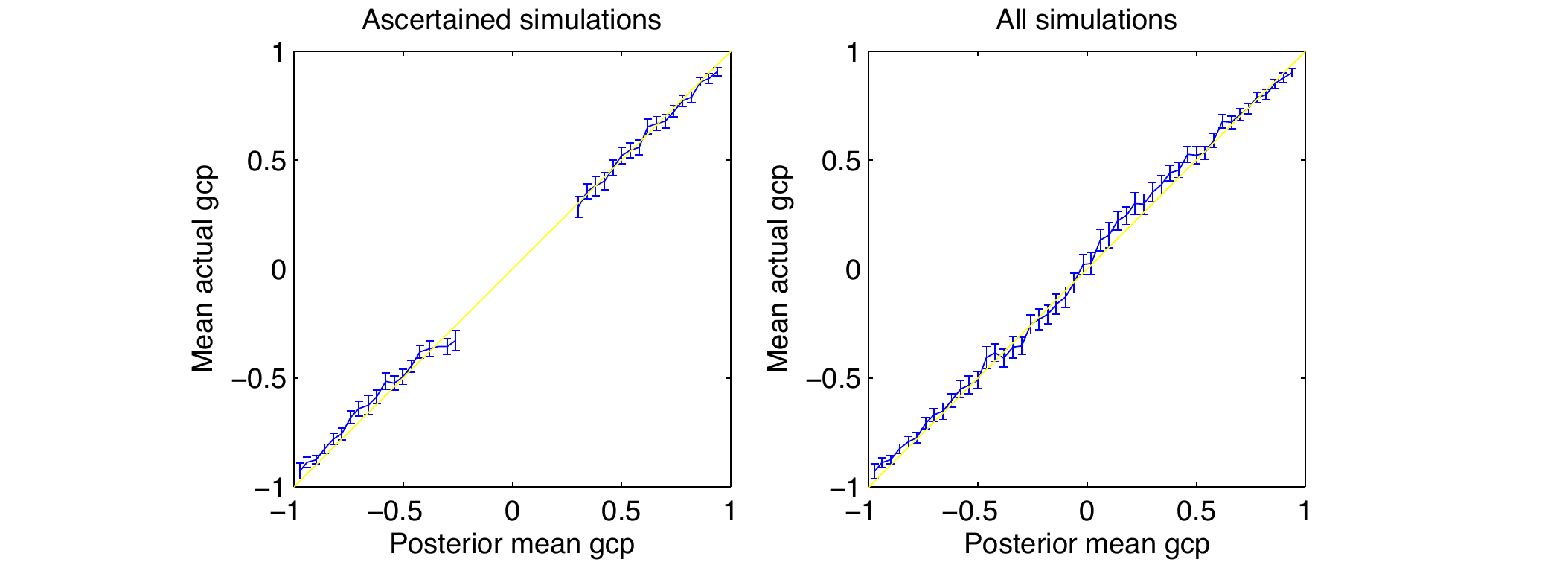}
\caption{Unbiasedness of posterior mean $\gcp$ estimates in simulations with LD and random true $\gcp$ values. Estimated values of $\gcp$ were binned and averaged, and mean true values of $\gcp$ are plotted for each bin, with standard errors. Points above the line indicate that $\gcp$ estimates were downwardly biased (toward -1). (a) Ascertained simulations ($43\%$) with significant genetic correlation ($p<0.05$) and evidence for partial causality ($p<0.001$). Only bins with count at least 10 are plotted. (b) All 10k simulations.}\label{bayes-fig}
\end{figure}

\subsection{Simulations with LD}

We performed simulations with LD to assess the robustness of LCV; we note that LD can potentially affect the performance of our method, which uses a modified version of LD score regression \cite{bulik1,bulik2} to normalize effect size estimates and to estimate genetic correlations. LD was computed using $M=596k$ common SNPs in $N=145k$ samples of European ancestry from the UK Biobank interim release \cite{sudlow}. Unlike our simulations with no LD, these simulations also included sample overlap. Because existing methods exhibited major limitations in simulations with no LD (Figure 2), we restricted these simulations to the LCV method.

First, we performed null simulations to assess calibration. We chose a set of default parameters similar to Figure~2b and varied each parameter in turn. In particular, similar to Figure~2, these simulations included uncorrelated pleiotropy, genetic correlations, differential polygenicity between the two traits, and differential power between the two traits  (Table~\ref{ld-supp-sims-table}a-m). LCV produced approximately well-calibrated or conservative false positive rates. Slight inflation was observed due to noise in our heritability estimates (Table~\ref{ld-supp-sims-table}c-m); proper calibration was restored by using constrained-intercept LD score regression\cite{bulik2} (resulting in more precise heritability estimates) (Table~\ref{fixedint-table}a-f). To avoid problems with noisy heritability estimates, we restrict our analyses of real traits to data sets with highly significant heritability estimates ($Z$ score for nonzero $h^2 = Z_h>7$).

Second, we performed causal simulations to assess power. We chose a set of default parameters similar to our null simulations, finding that LCV was well-powered (Table~\ref{ld-supp-sims-table}t), although its power was lower than in simulations with no LD (Figure~\label{MR-power-fig}a). We varied each parameter in turn, finding that power was reduced when we reduced the sample size, increased the polygenicity of the causal trait, reduced the causal effect size, or simulated a partially causal rather than fully causal genetic architecture (Table~\ref{ld-supp-sims-table}u-bb), similar to simulations with no LD (Figure~\label{MR-power-fig}b-f). These simulations indicate that LCV is well-powered to detect a causal effect for large GWAS under most realistic parameter settings, although its power does depend on genetic parameters that are difficult to predict.

Third, to assess the unbiasedness of $\gcp$ posterior mean (and variance) estimates, we performed simulations in which the true value of $\gcp$ was drawn uniformly from $[-1,1]$ (corresponding to the prior that LCV uses to compute its posterior mean estimates). We expected posterior-mean estimates to be unbiased in the Bayesian sense that $E(\gcp|\gcphat)=\gcphat$ (which differs from the usual definition of unbiasedness, that $E(\gcphat|\gcp)=\gcp$) \cite{goddard}. Thus, we binned these simulations by $\gcphat$ and plotted the mean value of $\gcp$ within each bin (Figure~\ref{bayes-fig}). We determined that mean $\gcp$ within each bin was concordant with $\gcphat$. In addition, the root mean squared error was 0.15, approximately consistent with the root mean posterior variance estimate of 0.13 (Table~\ref{bayes-table}).

In summary, we confirmed using simulations with LD that LCV produces well-calibrated false positive rates under a wide range of realistic genetic architectures; some p-value inflation was observed when heritability estimates were noisy, but false positives can be avoided in analyses of real traits by restricting to traits with highly significant heritability ($Z_h > 7$).  We also confirmed that LCV is well-powered to detect a causal effect under a wide range of realistic genetic architectures, and produces unbiased posterior mean estimates of the gcp.

\begin{table}[h]\small
\begin{tabular}{l  c l  c l  c l  c l  c l  c l  c l  c l  c l  c l  c l  c l  c l  c }
\hline
	& Regression coefficient (std err) & RMSE & RMPV \  \\ \hline
	Ascertained simulations (43\%) & 0.97 (.004) & 0.15 & 0.13 \\
	All simulations & 1.00 (.005) & 0.24 &  0.20 \\ 
	\\
	 \end{tabular}
\caption{Unbiasedness of estimated $\gcp$ and standard error in simulations with random true parameter values, using real LD.  We drew random values of $\gcp$ (and $\rho$) from a $Unif(-1,1)$ distribution and compared true and estimated values of $\gcp$, either for all 10k simulations or for a subset ($43\%$) of simulations in which the genetic correlation was nominally significant $p<0.05$ and the evidence for partial causality was strong ($p<0.001$).  We report the regression coefficient of true on estimated $\gcp$ values with standard error, as well as the root mean squared error and the root mean posterior variance estimate. }\label{bayes-table}
\end{table}

\subsection{Simulation details}
In order to simulate summary statistics with no LD, first, we chose causal effect sizes for each SNP on each trait according to the LCV model. The causal effect size vector for trait $k$ was 
\bal
\beta_k=\ff{h^2_k}{M}(q_k\pi+\gamma_k),
\end{align}
where in all simulations except for Figure~\ref{multi-intermediary-fig}, $q_k$ was a scalar, and $\pi$ and $\gamma_k$ were $1\times M$ vectors. In  Figure~\ref{multi-intermediary-fig}, $q_k$ was a $1\times 2$ vector and $\pi$ was a $2\times M$ matrix. Entries of $\pi$ were drawn from i.i.d. point-normal distribution with mean zero, variance 1, and expected proportion of causal SNPs equal to $p_\pi$. Entries of $\gamma_k$ were drawn from i.i.d. point-normal distributions with expected proportion of causal SNPs equal to $p_{\gamma_k}$; we modeled colocalization between non-mediated effects by fixing some expected proportion of SNPs $p_{\gamma_{1,2}}<\min(p_{\gamma_1},p_{\gamma_2})$ as having nonzero values of both $\gamma_1$ and $\gamma_2$. Then, we centered and re-scaled the nonzero entries of $\pi$ and $\gamma_k$, so that they had mean 0 and variance 1 and $1-q_k^2$, respectively. 

For simulations in  Figure~\ref{multi-intermediary-fig}, effect sizes were drawn from a mixture of Normal distributions: there was a point mass at (0,0); a component with $\ss_1=0,\ss_2\neq 0$; a component with $\ss_1\neq0,\ss_2=0$; and a component with $\ss_1\neq 0,\ss_2\neq 0,\sigma_{12}=\sqrt{\ss_1\ss_2}.$ 

Second, we simulated summary statistics as 
\bal \hat{\beta}_{k}{\sim} N(\beta_{k},\ff{1}{N_k}I),\end{align}
where $\beta_k$ is the vector of true causal effect sizes for trait $k$ and $N_k$ is the sample size for trait $k$. When we ran LCV on these summary statistics, we used constrained-intercept LD score regression rather than variable-intercept LD score regression both to normalize the effect estimates \cite{bulik2} and to estimate the genetic correlation \cite{bulik1}, with LD scores equal to one for every SNP.

In simulations with LD, we first simulated causal effect sizes for each trait in the same manner as simulations with no LD. Then, we obtained summary statistics in one of two ways, either using real genotypes or using real LD only. 

For other simulations, we simulated summary statistics without first simulating phenotypic values, using the fact that the sampling distribution of $Z$-scores is approximately \cite{conneely}:
\bal
Z{\sim} N(\sqrt{N}R\beta,R),
\end{align}
where $R$ is the LD matrix and $\beta$ is the vector of true effect sizes. We estimated $R$ from the $N=145k$ UK Biobank cohort using plink with an LD window size of 2Mb ($M=596k$), which we converted into a block diagonal matrix with 1001 blocks. The number 1001 was chosen instead of the number 1000 so that the boundaries of these blocks would not align with the boundaries of our 100 jackknife blocks; the use of blocks allowed us to avoid diagonalizing a matrix of size $596$k, while not significantly changing overall LD patterns (there are ${\sim} 50,000$ independent SNPs in the genome, and $1001<<50,000$). Because the use of a 2Mb window causes the estimated LD matrix to be non-positive semidefinite (even after converting it into a block diagonal matrix), each block was converted into a positive semidefinite matrix by diagonalizing it and removing its negative eigenvalues: that is, we replaced each block $A=V\Sigma V^T$ with the matrix $B$, where
$B=V \max(0,\Sigma) V^T.$
Then, because the removal of negative eigenvalues causes $B'$ to have entries slightly different from one, we re-normalized each block: $C=D^{-1/2}BD^{-1/2}$, where $D$ is the diagonal matrix corresponding to the diagonal of $B$. Even though the diagonal elements of $B$ are close to 1 (mostly between 0.99 and 1.01), this step is important to obtain reliable heritability estimates using LD score regression because otherwise the diagonal elements of the LD matrix will be strongly correlated with the LD scores ($r^2\approx 0.5$) and the heritability estimates will be upwardly biased, especially at low sample sizes.

We concatenated the blocks $C_1,...,C_{1001}$ to obtain a positive semi-definite block-diagonal matrix $R'$. We also computed and concatenated the matrix square root of each block. In order to obtain samples from a Normal distribution with mean $R'\beta$ and variance $\ff{1}{N}R'$, we multiplied a vector having independent standard normal entries by the matrix square root of $R'$ and added this noise vector to the vector of true marginal effect sizes, $R'\beta$. We computed LD scores directly from $R$. For simulations with sample overlap, the summary statistics were correlated between the two GWAS: the correlation between the noise term in the estimated effect of SNP $i$ on trait~1 and the estimated effect of SNP $j$ on trait~2 was $R_{ij}'\rho_\text{total}N_\text{shared}/\sqrt{N_1N_2}$, which is the amount of correlation that would be expected if the total (genetic plus environmental) correlation between the traits is $\rho_\text{total}$ \cite{bulik1}.
\section{Discussion of additional trait pairs}

We briefly discuss several other trait pairs with significant evidence of partial genetic causality, including novel results and results that have previously been reported (Table~\ref{all-sig-table}).
\begin{itemize}

\item We identified four traits with evidence for a fully or partially genetically causal effect on hypertension (Table~\ref{all-sig-table}), which is genetically correlated with MI ($\hat{\rho}_g=0.49(0.10)$). These included genetically causal effects of BMI, consistent with the published literature \cite{pickrell,lyall}, as well as triglycerides and HDL. The genetically causal effect of HDL indicates that there exist major metabolic pathways affecting hypertension with little or no corresponding effect on MI. The positive partially genetically causal effect of reticulocyte count, which had a low gcp estimate ($\gcphat=0.41(0.13)$), is likely related to the substantial genetic correlation of reticulocyte count with triglycerides ($\hat{\rho_g}=0.33(0.05)$) and BMI ($\hat{\rho_g}=0.39(0.03)$).
\item We detected evidence for a fully or partially genetically causal effect of triglycerides on five cell blood traits: mean cell volume, platelet distribution width, reticulocyte count, eosinophil count  and monocyte count (Table~1). These results highlight the pervasive effects of metabolic pathways, which can induce genetic correlations with cardiovascular phenotypes. For example, shared metabolic pathways may explain the high genetic correlation of reticulocyte count with MI ($\hat{\rho}_g=0.31(0.06)$) and hypertension ($\hat{\rho}_g=0.27(0.04)$).
\item There was evidence for a negative fully or partially genetically causal effect of balding on number of children in males. Two possible explanations are shared pathways involving androgens \cite{ellis} and sexual selection against early balding.
\item There was evidence for a positive fully or partially genetically causal effect of BMI on triglycerides, consistent with results using MR \cite{lyall} and bidirectional MR \cite{pickrell}. There was also evidence for a positive genetically causal effect of LDL on the self-reported high cholesterol phenotype, consistent with LDL cholesterol representing one component of this compound phenotype.
\item There was evidence for fully or partially genetically causal effects of several traits on various platelet phenotypes: large negative effects on platelet count for platelet distribution width and platelet volume, and effects of triglycerides and HDL on platelet distribution width.
\item It has been suggested that height has a causal effect on educational attainment \cite{tyrell}. While our results support a partially genetically causal effect, the low gcp estimate ($\gcphat=0.33(0.10)$) suggests shared developmental pathways rather than direct causality, highlighting the benefit of our non dichotomous approach to causal inference. There was a similar result for age at menarche and height, which was previously reported using Bidirectional MR \cite{pickrell}.
\item A recent study reported genetic correlations between various complex traits and number of children in males and females\cite{sanjak}. We identified only one trait (balding in males) with a fully or partially causal effect on number of children (negative effect on number of children in males). (Two possible explanations are shared pathways involving androgens \cite{ellis} and sexual selection against early balding.) For college education, which has a strong negative genetic correlation with number of children ($\hat{\rho_g}=-0.31(0.07)$ and $-0.26(0.06)$ in males and females respectively), we obtained low gcp estimates with low standard errors ($\gcphat=0.00(0.09)$ and $\gcphat=0.04(0.21)$ respectively), providing evidence against causality. Thus, a genetic correlation with number of children does not imply a causal effect. This result does not contradict the conclusion of reference \citen{sanjak} that complex traits are under natural selection, as natural selection produces a change in the mean value of a trait even if the trait is non-causally correlated with fitness \cite{gprice}.
\end{itemize}

\end{document}